\documentclass{comsoc2020}
\usepackage{times}
\usepackage{graphicx}
\usepackage{latexsym}
\usepackage{tikz}
\usepackage{float}

\usepackage{xcolor}
\usepackage{amsthm}
\usepackage{amsmath,amsfonts}

\usepackage{thmtools}

\usepackage[utf8]{inputenc}  
\usepackage[T1]{fontenc}   
 
\newcommand{\nbCandidates}{\emph{m} }
\newcommand{\nbVoters}{\emph{n} }

\theoremstyle{definition}
\declaretheorem[name=Example,qed={\lower-0.3ex\hbox{$\triangle$}}]{ex}
\newtheorem{definition}{Definition}[section]
\newtheorem{thm}{Theorem}[section]
\newtheorem{prop}{Proposition}[section]

\title{Recognizing Single-Peaked Preferences\\ on an Arbitrary Graph: Complexity and Algorithms}
\author{Bruno Escoffier and Olivier Spanjaard and Magdaléna Tydrichov\'a}

\begin{document}


\begin{abstract}
  This paper is devoted to a study of single-peakedness on arbitrary graphs. Given a collection of preferences (rankings of a set of alternatives), we aim at determining a connected graph $G$ on which the preferences are single-peaked, in the sense that all the preferences are traversals of $G$. Note that a collection of preferences is always single-peaked on the complete graph. We propose an Integer Linear Programming formulation (ILP) of the problem of minimizing the number of edges in $G$ or the maximum degree of a vertex in $G$. We prove that both problems are NP-hard in the general case. However, we show that if the optimal number of edges is $\nbCandidates\!-\!1$ (where $\nbCandidates$ is the number of candidates) then any optimal solution of the ILP is integer and thus the integrality constraints can be relaxed. This provides an alternative proof of the polynomial time complexity of recognizing single-peaked preferences on a tree. We prove the same result for the case of a path (an axis), providing here also an alternative proof of polynomiality of the recognition problem. Furthermore, we provide a polynomial time procedure to recognize single-peaked preferences on a pseudotree (a connected graph that contains at most one cycle). We also give some experimental results, both on real and synthetic datasets.
\end{abstract}


\section{INTRODUCTION}

Aggregating the preferences of multiple agents is a primary task in many applications of artificial intelligence, e.g., in preference learning \cite{chengHullermeier2009,ClemenconKS18} or in recommender systems \cite{andersen2008trust,pennock2000social}. The individual preferences of the agents are often represented as rankings on a set of alternatives, where the alternatives may be cultural products (books, songs, movies...), technological products, candidates for an election, etc. The aim of preference aggregation is then to produce an aggregate ranking from a collection of rankings (called preference profile).

The preferences are said to be \emph{structured} if they share some common structure \cite{elkind2017structured}. For example, in a political context, it is conventional to assume that each individual preference is decreasing as one moves away from the preferred candidate along a left-right axis on the candidates, axis on which individuals all agree. Such preferences are called \emph{single-peaked} \cite{black1948rationale}.
They have been the subject of much work in social choice theory. The most well-known result states that if preferences are single-peaked, then one escapes from Arrow's impossibility theorem. We recall that Arrow's theorem states that any unanimous aggregation function for which the pairwise comparison between two alternatives is independent of irrelevant alternatives is dictatorial. Furthermore, from the computational viewpoint, many NP-hard social choice problems (e.g., Kemeny rule and Young rule for rank aggregation \cite{brandt2015bypassing}, Chamberlin-Courant rule for proportional representation \cite{betzler2013computation}) become polynomially solvable if the preferences are single-peaked. 

Given the axiomatic and algorithmic consequences, the question of the computational complexity of recognizing single-peaked preferences is thus natural. Bartholdi and Trick \cite{bartholdi1986stable} have proposed an $O(nm^2)$ algorithm to compute a compact representation of \emph{all} axes on which a collection of $n$ preferences on $m$ candidates are single-peaked, or state that none exists. This complexity can be decreased to $O(nm)$ if one looks for only \emph{one} possible axis \cite{escoffier2008single}.

Several classes of structured preferences have been proposed in the literature in order to generalize the single-peaked domain with respect to an axis, i.e., a path, to more general graphs. Given a set $\mathcal{C}\!=\! \{1,\ldots,m\}$ of candidates, a preference order $\succ$ over $\mathcal{C}$ is single-peaked on an undirected graph $\mathcal{G}\!=\! (\mathcal{C},\mathcal{E})$ if it is a \emph{traversal} of $\mathcal{G}$, i.e., for each $j\!\in\!\mathcal{C}$ the upper-contour set $\{i \!\in\!\mathcal{C} : i\!\succ\!j\}$ is connected. A preference profile is then single-peaked on $\mathcal{G}$ if every preference is single-peaked on $\mathcal{G}$. Demange studied single-peakedness on a tree \cite{demange1982single}; Peters and Lackner studied single-peakedness on a circle~\cite{peters17preferences}. 

Some good axiomatic properties remain valid when preferences are single-peaked on a tree: if the number of voters is odd, such profiles still admit a Condorcet winner (a candidate who is preferred over each other candidate by a majority of voters) \cite{demange1982single}, and returning this Condorcet winner is a strategyproof voting rule. On the contrary, every majority relation can be realized by a collection of preferences that are single-peaked on a circle \cite{peters17preferences}, and consequently single-peaked preferences on a circle do not inherit the good axiomatic properties of single-peakedness on an axis regarding voting rules that are based on the majority relation.


The goal of this paper is to study the recognition problem for single-peaked preferences on arbitrary connected graphs.
Although one cannot expect social choice theoretic guarantees from single-peakedness on an arbitrary graph (as it does not result in a domain restriction, in the sense that any preference profile is single-peaked on an arbitrary graph), the knowledge of such graphs indeed provides information based on preferences that gives some insights on the similarity of candidates/items, that could be used for instance in recommendation systems. For instance, assume that one discovers that the preferences over movies $\{1,2,3,4,5\}$ are single-peaked w.r.t. an axis (1,2,3,4,5) over the movies. If ones knows that an agent likes movies 3 and 5, then it is natural to recommend to watch movie 4. More generally, one can take advantage of single-peakedness on a sparse graph in order to make recommendations in the neighbourhood of liked items.
Thereby, we focus here on determining a graph that minimizes (1) the number of edges of the graph or (2) the maximum degree of a vertex.
This choice is motivated by the fact that
these criteria are measures of sparsity of a graph (the sparsest the graph is, the more informative), but also because they generalize known cases such as paths, cycles and trees. 
Let us indeed emphasize that the mathematical programming approach we propose to identify a graph generalizes the best known instances of the single-peaked recognition problem and provides a uniform treatment of them, leading to simple polynomial time algorithms.

\medskip

\noindent \textbf{Our contribution.}  We propose here an Integer Linear Programming formulation (ILP) of problems (1) and (2), and we show that both of them are NP-hard. Nevertheless, if the optimal value for problem (1) is $m\!-\!1$ (where $m$ is the number of candidates), we prove the integrality of the optimal basis solution of the linear program obtained by relaxing the integrality constraint in the ILP. This provides an alternative polynomial time method, based on a simple linear programming solver, to recognize single-peakedness on a tree, as a connected graph with $m$ vertices and $m\!-\!1$ edges is a tree. By adding some constraints on the max degree of a vertex, we obtain the same result for the case of paths. As a last theoretical result, we prove that preferences single-peaked on a \emph{pseudotree} (a connected graph that contains at most one cycle) can be recognized in polynomial time. We also provide some experimental results, both on real-world and synthetic datasets, where we measure the density of the graphs depending on the diversity of preferences of voters.

\medskip

\noindent\textbf{Related work.} We briefly describe here some previous contributions that have addressed the concept of single-peakedness on arbitrary graphs, the optimization view of the recognition problem and the use of integer linear programming formulations for computational social choice problems related to structured preferences:\\ [0.5ex]  
$\bullet$~~Nehring and Puppe defined a general notion of single-peaked preferences based on abstract betweenness relations between candidates \cite{nehring2007structure}. In their setting, it is possible to define single-peaked preferences on a graph $G$ by considering the \emph{graphic betweenness relation}: candidate $j$ is between candidates $i$ and $k$ if and only if $j$ lies on a shortest path between $i$ and $k$ in $G$. A preference profile is then single-peaked on $G$ if for every preference $\succ$, if $i^*$ is the most preferred candidate w.r.t. $\succ$ and $j$ is on a shortest path between $i^*$ and $k$ then $j\!\succ\! k$. This definition enables them to state general results regarding strategyproofness on restricted domains of preferences. Note that this definition of single-peakedness on a graph does not coincide with the one we use.\\ [0.5ex] 
$\bullet$~~Peters and Elkind showed how to compute in polynomial time a compact representation of \emph{all} trees with respect to which a given profile is single-peaked \cite{peters2016preferences}. This structure allows them to find in polynomial time trees that have, e.g., the minimum
degree, diameter, or number of internal nodes among all trees
with respect to which a given profile is single-peaked. We provide here alternative proofs for some of these results, based on linear programming arguments.\\ [0.5ex]
$\bullet$ Peters recently proposed ILP formulations for proportional representation problems, and showed that the binary constraint matrix is totally unimodular if preferences are single-peaked, because the matrix has then the consecutive ones property \cite{peters18single}. We recall that the vertices of a polyhedron defined by a totally unimodular constraint matrix are all integer, thus solving the linear programming relaxation yields an optimal solution to the original ILP problem. We also rely on linear programming for proving the polynomial time complexity of some of the recognition problems we tackle here. 


\medskip

\noindent\textbf{Organization of the paper.} The two optimization variants of the recognition problem tackled in the paper are defined in Section~\ref{sec:ILPandComplexity}. The two problems are proved NP-hard, and an ILP formulation is proposed. In Section~\ref{sec:lp}, we consider the continuous relaxation of the ILP to show how to recognize single-peaked preferences on a path or a tree by linear programming, thus providing a unified view of known polynomiality results. Section~\ref{sec:Pseudotrees} is devoted to a new tractable case of recognition problem, namely the recognition of single-peaked preferences on a pseudotree. Finally, the results of numerical tests are presented in Section~\ref{sec:ExperimentalStudy}, on both real and synthetic data, to give some hints about the kind of graphs that are returned by the ILP on real data, and to study how the density of the graph varies with the number of voters and the diversity of preferences.

\section{ILP FORMULATION AND COMPLEXITY}
\label{sec:ILPandComplexity}

\subsection{Problem Definition}
We start by recalling some basic terminology of social choice theory. 
Given a set $\mathcal{C}\!=\!\lbrace 1, 2 , \ldots, \nbCandidates \rbrace$ of  candidates and a set $\{v_1, \ldots, v_\nbVoters\}$ of voters, each voter $v_i$ ranks all candidates from the most to the least preferred one. This ranking is called the \emph{preference} of $v_i$. It is simply a permutation of $\mathcal{C}$, which can be formally described as an \nbCandidates-tuple $R_i\!=\!(\pi_i(1), \ldots , \pi_i(\nbCandidates))$, where $\pi_i(k)$ is the \emph{k}-th most preferred candidate of voter $v_i$. The set $\mathcal{P} = \lbrace R_1, \ldots , R_\nbVoters \rbrace $ of preferences of all voters is called the \emph{profile}.   

As emphasized in the introduction, several definitions of single-peakedness on an arbitrary graph can be found in the literature. In our study, we are using the following one \cite{elkind2017structured}: 

\begin{definition}{\textbf{Single-peakedness on arbitrary graph (SP)}} Let $\mathcal{C}$ be a set of \nbCandidates candidates and $\mathcal{P}$ the profile of preferences of \nbVoters voters. Let $\mathcal{G}\!=\!(\mathcal{C}, E)$ be a connected undirected graph. We say that $\mathcal{P}$ is \emph{single-peaked on the graph} $\mathcal{G}$ (SP) if every $R_i\!\in\!\mathcal{P}$ is a traversal of $\mathcal{G}$, i.e., for each $R_i\!\in\! \mathcal{P}$ and for each $k\!\in\!\lbrace 1, \ldots \nbCandidates \rbrace$, the subgraph of $\mathcal{G}$ induced by the vertices $\lbrace \pi_i(1), \ldots , \pi_i(k) \rbrace$  is connected.
\end{definition}

This notion of single-peakedness coincides with the standard definition on an axis/cycle/tree when $\mathcal{G}$ is a path/cycle/tree. In this article, when a profile $\mathcal{P}$ is single-peaked w.r.t. a graph $\mathcal{G}$, for conciseness we will say that $\mathcal{P}$ is \emph{compatible} with $\mathcal{G}$ (or that $\mathcal{G}$ is compatible with $\mathcal{P}$).

\begin{ex}\label{ex21}
Consider the profile with 4 voters and 5 candidates: 
\begin{align*}
R_1 : & \ (1, 2, 3, 4,5) \\
R_2 : & \ (1, 3, 4, 2,5) \\
R_3 : & \ (2, 5, 3, 4,1) \\
R_4 : & \ (3, 5, 4, 2,1) 
\end{align*}
Note that, for $R_1$, the connectivity constraint applied to the first two candidates makes the edge $\{1,2\}$ necessary in the graph. The same occurs for $\{1,3\}$, $\{2,5\}$ and $\{3,5\}$. Thus, any solution contains the 4-cycle $(1,2,5,3,1)$ (in particular the profile is not SP on a tree or on a cycle). One can easily check that adding edge $\{3,4\}$ makes a graph with 5 edges compatible with the profile, and this is the (unique) optimal solution if we want to minimize the number of edges. 
\end{ex}


    
Obviously, any profile is SP on the complete graph. However, this case is not interesting because it does not give any information about the preference structure. That is why we are looking for a \emph{minimal} graph on which the profile is SP. The notion of minimality needs to be made more precise. In our study, we focus essentially on minimizing the number of graph edges. Another criterion we consider is the minimization of the maximum degree of vertices. 
Put another way, given a preference profile $\mathcal{P}$, we want to determine a graph $\mathcal{G}$ on which the profile is SP, so as to minimize either the number of edges of $\mathcal{G}$, or its (maximum) degree. We emphasize the fact that:
\begin{itemize}
    \item minimizing the number of edges allows to detect when the profile is compatible with a tree (this occurs iff the minimum number of edges is $\nbCandidates\!-\!1$, since $\mathcal{G}$ is necessarily connected);
    \item minimizing the degree of $\mathcal{G}$ allows to detect when the profile is compatible with a cycle (this occurs iff there exists a graph $\mathcal{G}$ with maximum degree 2);
    \item combining the objective allows to detect when the profile is compatible with an axis: this occurs iff there is a graph $\mathcal{G}$ with maximum degree 2 and $\nbCandidates\!-\!1$ edges.
\end{itemize}
So the tackled problems generalize the most well known (tractable) recognition problems of single-peakedness.

\subsection{ILP Formulation}\label{sec:ILP}
We now present an ILP formulation of the tackled problems. We are looking for a graph $\mathcal{G}$ with \nbCandidates vertices. For each pair  $\lbrace k, l \rbrace\!\subseteq\!\lbrace 1, \ldots , \nbCandidates \rbrace$ of vertices, we define a binary variable $x_{\lbrace k, l \rbrace }$ such that 
\begin{equation*}
 x_{\lbrace k, l \rbrace} = \begin{cases} 1 & \text{if edge $\{k,l\}$ is present in graph $\mathcal{G}$,} \\
 0 & \text{otherwise.}
 \end{cases}
\end{equation*}
Hence, if we are minimizing the number of graph edges, the objective function $f(x)$ takes the form 
$$ f(x)=\textstyle\sum_{\lbrace k, l \rbrace \subseteq \lbrace 1, \ldots, \nbCandidates \rbrace} x_{\lbrace k, l \rbrace}.$$
If the objective is to minimize the maximum degree, then 
$$f(x) = \max_{k \in \lbrace 1, \hdots , \nbCandidates \rbrace } \sum_{l =1, l\neq k}^{\nbCandidates} x_{\lbrace k, l \rbrace }.$$
In this latter case, the classical way of linearizing $f(x)$ in a minimization setting is to introduce an auxiliary variable $z$ as follows:
\begin{align*}
& \min z \\
& \text{s.t. } \sum_{l =1, l\neq k}^{\nbCandidates} x_{\lbrace k, l \rbrace } \leq z \ \  \forall k\!\in\!\lbrace 1, \hdots , \nbCandidates \rbrace
\end{align*}

Regardless of the objective function, the other constraints of the problem remain the same. Each $R_i\!=\!(\pi_i(1), \ldots , \pi_i(\nbCandidates))$ for $i\!\in\! \lbrace 1, \ldots , \nbVoters \rbrace$ has to be a graph traversal. In other words, for each $k \in \lbrace 2, \ldots \nbCandidates \rbrace$, $\pi_i(k)$ is connected to at least one of the vertices $\pi_i(1), \ldots, \pi_i(k-1)$. In terms of LP constraints, it is formulated as follows: 
\begin{align*}
\sum_{j = 1}^{k-1} x_{\{\pi_i(j), \pi_i(k)\}} \geq 1 \ \ \forall i \in \lbrace 1, \hdots, \nbVoters \rbrace, \forall k \in  \lbrace 2, \hdots, \nbCandidates \rbrace.
\end{align*}
To sum up, the ILP formulation of the tackled problems is:
\begin{align*}
 & \min f(x) \\
\mbox{s.t.} \ & \begin{cases}
\sum_{j = 1}^{k-1} x_{\{\pi_i(j), \pi_i(k)\}} \geq 1 \ \ \forall i \in \lbrace 1, \hdots, \nbVoters \rbrace, \forall k \in  \lbrace 2, \hdots, \nbCandidates \rbrace \\
x_{\lbrace k, l \rbrace}\!\in\! \lbrace 0, 1 \rbrace \ \forall \lbrace k, l \rbrace \subseteq \lbrace 1, \hdots , \nbCandidates \rbrace
\end{cases}
\end{align*}
\begin{ex}
Consider profile $\mathcal{P}\!=\!\{(1,4,2,3),(3,4,2,1)\}$. The ILP formulation of the problem of determining a graph $\mathcal{G}$ compatible with $\mathcal{P}$ while minimizing the number of graph edges is: 
\begin{align*}
\min \ & x_{\lbrace 1, 2 \rbrace} + x_{\lbrace 1, 3 \rbrace} + x_{\lbrace 1, 4 \rbrace} + x_{\lbrace 2, 3 \rbrace} + x_{\lbrace 2, 4 \rbrace} + x_{\lbrace 3, 4 \rbrace} \\ 
\mbox{s.t.} & \begin{cases}
x_{\lbrace 1,4 \rbrace} \geq 1 \\
x_{\lbrace 1,2 \rbrace} + x_{\lbrace 2,4 \rbrace } \geq 1 \\
x_{\lbrace 1,3 \rbrace} + x_{\lbrace 3,4 \rbrace } + x_{\lbrace 2,3 \rbrace}\geq 1 \\
\\
x_{\lbrace 3,4 \rbrace} \geq 1 \\
x_{\lbrace 2,3 \rbrace} + x_{\lbrace 2,4 \rbrace } \geq 1 \\
x_{\lbrace 1,3 \rbrace} + x_{\lbrace 1,4 \rbrace } + x_{\lbrace 1,2 \rbrace}\geq 1 \\
\\
x_{\lbrace k, l \rbrace}\!\in\! \lbrace 0, 1 \rbrace \ \forall \lbrace k, l \rbrace\!\subset\!\lbrace 1, 2, 3, 4 \rbrace
\end{cases}
\end{align*}
\end{ex}

\subsection{Minimizing the Number of Edges}
In this section, we study the computational complexity of the problem of  minimizing the number of edges of $\mathcal{G}$. As a first observation, note that we cannot use the continuous relaxation of this ILP  ($x_{\{k,l\}}\!\in\![0,1]$) to solve the problem. The following example indeed shows that the optimal solution (when minimizing the number of edges) of this relaxation is not necessarily integer:
\begin{ex}
Consider the profile with 3 voters and 4 candidates:
\begin{align*}
R_1 : & \ (1,2,4,3) \\
R_2 : & \ (2,3,4,1) \\ 
R_3 : & \ (1,3,4,2) 
\end{align*}
From the two first options of each voter, we see immediately that the edges $\lbrace 1,2 \rbrace$, $\lbrace 2,3 \rbrace $ and $\lbrace 1,3 \rbrace$ are necessarily present in the graph. Then, we observe that vertex $4$ needs to be connected to at least one of vertices $1$ and $2$, at least one of vertices $2$ and $3$ and finally at least one of  vertices $1$ and $3$. Consequently any integer solution of the problem will be a graph with at least 5 edges. However, there exists a fractional solution of the continuous relaxation with value 4.5: 
\begin{center}
\begin{tikzpicture}[scale=0.7]
    \node[shape=circle,draw=black] (1) at (0,3) {1};
    \node[shape=circle,draw=black] (2) at (3,3) {2};
    \node[shape=circle,draw=black] (3) at (3,0) {3};
    \node[shape=circle,draw=black] (4) at (0,0) {4};

    \path [-] (1) edge node[pos=0.5,above] {1} (2);
    \path [-](1) edge node[pos=0.2, right] {1} (3);
    \path [-] (3) edge node[pos=0.2, right] {1} (2);
    
    \path [dashed] (1) edge node[pos =0.5, left]{0.5}(4);
    \path [dashed] (2) edge node [pos=0.2,left]{0.5}(4);
    \path [dashed] (3) edge node [pos=0.5,above]{0.5}(4);

\end{tikzpicture}
\end{center}
\end{ex}

We now show that the problem is actually NP-hard. 



\begin{thm}
Given a preference profile $\mathcal{P}$, it is NP-hard to find a graph compatible with $\mathcal{P}$  with a minimum number of edges. 
\end{thm}
\proof
We use a polynomial time reduction from the set cover problem, known to be NP-hard \cite{Garey79}. We recall its definition:\\ [1.5ex]
\noindent \emph{Set cover problem:} Given a finite set $\mathcal{U}\!=\!\lbrace e_1, \ldots e_\nbVoters \rbrace$  of elements,  a set $\mathcal{S}\!=\!\lbrace S_1, \ldots , S_\nbCandidates \rbrace$ of subsets  of $\mathcal{U}$ and $k\!\in\!\mathbb{N}$, the question is to determine if there exists a  subset $\mathcal{C}\!\subseteq\!\mathcal{S}$ of size $k$ such that $\displaystyle\cup_{S \in  \mathcal{C}} S\!=\!\mathcal{U}$.\\ [1.5ex] 
From an instance of the set cover problem, we define a preference profile $\mathcal{P}$ as follows:
\begin{description}
    \item[$(i)$] Let $\lbrace S_1, \ldots, S_\nbCandidates, z \rbrace$ be a set of candidates.
    \item[$(ii)$] Let $\lbrace v_1, \ldots , v_\nbVoters\rbrace$ be a set of voters. Let $S_{i_1}, \ldots S_{i_l}$ be the subsets in $\mathcal{S}$ containing element $e_i\!\in\! \mathcal{U}$, and $S_{i_{l+1}}, \ldots , S_{i_\nbCandidates}$ the other subsets in $S$. Then, the preference of voter $v_i$ is defined as 
    $$ R_i = (S_{i_1}, \ldots , S_{i_l} , z , S_{i_{l+1}}, \ldots , S_{i_\nbCandidates}).$$
    \item[$(iii)$] We add $\frac{\nbCandidates \cdot (\nbCandidates - 1)}{2}$ voters $v_{\lbrace i,j \rbrace} , \lbrace i,j \rbrace\!\subseteq\! \lbrace 1, \ldots , \nbCandidates\}$ such that 
    $$ R_{\lbrace i, j \rbrace} = (S_i, S_j, \underbrace{S_1, \hdots , S_\nbCandidates}_{\text{except } S_i, S_j}, z).$$
\end{description}
We prove that there exists a set cover of size $k$  if and only if there exists a graph $\mathcal{G}$ compatible with $\mathcal{P}$ that has $\frac{\nbCandidates \cdot (\nbCandidates - 1)}{2} + k$ edges.

Let $\mathcal{C}$ be a set cover solution of size $k$. We generate a graph $\mathcal{G}$ compatible with $\mathcal{P}$ in the following manner:
\begin{itemize}
    \item[$a)$] For each $\lbrace i, j \rbrace\!\in\! \lbrace 1, \ldots , \nbCandidates \rbrace ^2, i\!\neq\!j$, the edge $\lbrace S_i, S_j \rbrace$ is in $\mathcal{G}$ - this is necessary for the preferences of type $(iii)$ above to be SP on $\mathcal{G}$. 
    \item[$b)$] For each $i\!\in\!\lbrace 1, \ldots , \nbCandidates \rbrace$, the edge $\lbrace S_i, z \rbrace$ is in $\mathcal{G}$ if and only if $S_i\!\in\!\mathcal{C} $.
\end{itemize}
Hence, the subgraph formed by vertices $\lbrace S_1, \ldots , S_\nbCandidates \rbrace$ is a clique having $\frac{\nbCandidates \cdot (\nbCandidates - 1)}{2}$ edges, and there are exactly $k$ more edges adjacent to $z$ - in total, $\mathcal{G}$ has $\frac{\nbCandidates \cdot (\nbCandidates - 1)}{2}\!+\!k$ edges. As $k\!>\!0$, the graph is connected and all preferences of type $(iii)$ are SP on $\mathcal{G}$. Let $R_i$ be one of the preferences of type $(ii)$. We need to prove that $z$ is connected to at least one of the vertices $S_{i_1}, \ldots , S_{i_l}$. As the sets $S_{i_1}, \ldots , S_{i_l}$ are the only sets of $\mathcal{S}$ containing the element $e_i$, and as $\mathcal{C}$ is a solution of the set cover instance, this is true due to $b)$. So, $\mathcal{G}$ is a graph compatible with $\mathcal{P}$ that has $\frac{\nbCandidates \cdot (\nbCandidates - 1)}{2} \!+\!k$ edges.

To prove the other implication, let $\mathcal{G}$ be a graph compatible with $\mathcal{P}$ that has $\frac{\nbCandidates \cdot (\nbCandidates - 1)}{2} + k$ edges. As $\mathcal{G}$ is compatible with $\mathcal{P}$, the subgraph induced by the set of vertices $\lbrace S_1, \ldots S_\nbCandidates \rbrace$ must be a clique so that the preferences $R_{\lbrace i, j \rbrace}$ of type $(iii)$ are SP on $\mathcal{G}$. Hence, this subgraph contains $\frac{\nbCandidates \cdot (\nbCandidates - 1)}{2}$ edges, and so, there are exactly $k$ edges adjacent to $z$.  Let us define $\mathcal{C}$ containing $S_i$ iff $S_i$ is adjacent to $z$ in $\mathcal{G}$.
As $\mathcal{G}$ is compatible with $\mathcal{P}$, each preference $R_i$ of type $(ii)$ is SP on $\mathcal{G}$. It means that at least one of $S_{i_1}, \ldots , S_{i_l}$ is adjacent to $z$, so is in $\mathcal{C}$. As all these sets contains $e_i$, there is an element of $\mathcal{C}$ that covers $e_i$. The subset $\mathcal{C}\!\subseteq\! \mathcal{S}$ is thus a solution of size $k$ of the set cover instance. 
\endproof

\subsection{Minimizing the Maximum Degree}

We now consider our second objective function, namely the maximum degree of a vertex in the graph (to be minimized). We come up with similar results.

First, as for the minimization of the number of edges, the ILP formulation we have proposed in Section~\ref{sec:ILP} is not integer, as we can see in the following case:
\begin{ex}
Consider a profile with 3 candidates and one voter with ranking $(1,2,3)$. The ILP formulation of the problem of determining a graph $\mathcal{G}$ of minimum degree compatible with $\mathcal{P}$: 
\begin{align*}
\min & \ z \\
\mbox{s.t.} & \begin{cases}
x_{\lbrace 1,2 \rbrace} + x_{\lbrace 1,3 \rbrace} \leq z \\
x_{\lbrace 1,2 \rbrace} + x_{\lbrace 2,3 \rbrace} \leq z \\
x_{\lbrace 1,3 \rbrace} + x_{\lbrace 2,3 \rbrace} \leq z \\
\\
x_{\lbrace 1,2 \rbrace} \geq 1 \\
x_{\lbrace 1,3 \rbrace} + x_{\lbrace 2,3 \rbrace} \geq 1
\\
x_{\lbrace k, l \rbrace}\!\in\! \lbrace 0, 1 \rbrace \ \forall \lbrace k, l \rbrace\!\subset\!\lbrace 1, 2, 3 \rbrace
\end{cases}
\end{align*}
The value of an optimal integer solution is $z\!=\!2$, but there exists a fractional solution of the continuous relaxation with value 1.5: 
\begin{center}
\begin{tikzpicture}[scale=0.7]
    \node[shape=circle,draw=black] (1) at (0,3) {1};
    \node[shape=circle,draw=black] (2) at (3,3) {2};
    \node[shape=circle,draw=black] (3) at (3,0) {3};

    \path [-] (1) edge node[pos=0.5,above]{1} (2);
    \path [dashed] (1) edge node[pos =0.5, left]{0.5}(3);
    \path [dashed] (2) edge node[pos =0.5, left]{0.5}(3);

\end{tikzpicture}
\end{center}
\end{ex}

Here again, we show that the problem of minimizing the degree of $\mathcal{G}$ is NP-hard, by a similar reduction sketched below.

\begin{thm}
Given a preference profile $\mathcal{P}$, it is NP-hard to find a graph compatible with $\mathcal{P}$  with a minimum degree. 
\end{thm}
\begin{proof}
 Let $\mathcal{U}\!=\!\lbrace e_1, \ldots e_\nbVoters \rbrace$, $\mathcal{S}\!=\! \lbrace S_1, \ldots , S_\nbCandidates \rbrace$, $k\!\in\!\mathbb{N}$ be an instance of the set cover problem. Consider the profile $\mathcal{P}$ defined as follows:
\begin{description}
    \item[$(i)$] Let $\lbrace S_1, \ldots, S_\nbCandidates, z, t_1,\dots,t_\nbCandidates \rbrace$ be a set of candidates.
    \item[$(ii)$] Let $\lbrace v_1, \ldots , v_\nbVoters\rbrace$ be a set of voters. Let $S_{i_1}, \ldots S_{i_l}$ be the subsets in $\mathcal{S}$ containing the element $e_i\!\in\! \mathcal{U}$, and $S_{i_{l+1}}, \ldots , S_{i_\nbCandidates}$ the other subsets in $S$. Then, the preference of voter $v_i$ is defined as 
    $$ R_i = (S_{i_1}, \ldots , S_{i_l} , z , S_{i_{l+1}}, \ldots , S_{i_\nbCandidates},t_1,\dots,t_\nbCandidates).$$
    \item[$(iii)$] We add $\frac{\nbCandidates \cdot (\nbCandidates - 1)}{2}$ voters $v_{\lbrace i,j \rbrace} , \lbrace i,j \rbrace \!\subseteq\! \lbrace 1, \ldots , \nbCandidates\}$ such that 
    $$ R_{\lbrace i, j \rbrace} = (S_i, S_j, \underbrace{S_1, \hdots , S_\nbCandidates}_{\text{except } S_i, S_j}, z,t_1,\dots,t_\nbCandidates).$$
    \item[$(iv)$] We add $\nbCandidates$ voters $v^{t_1},\dots,v^{t_\nbCandidates}$ where the preference of $v^{t_i}$ is defined as
    $$R^{t_i}=(z,t_i, \underbrace{t_1, \hdots , t_\nbCandidates}_{\text{except } t_i}, S_1,\dots,S_\nbCandidates),$$
    and a voter $w$ with preference $(t_1,S_1,S_2,\dots,S_\nbCandidates, z, t_2,\dots,t_\nbCandidates)$.
\end{description}

Note that in any graph $\mathcal{G}$ compatible with the profile:
\begin{itemize}
\item[$a)$] the $S_i$ form a clique (edge $\{S_i,S_j\}$ is enforced by voter $v_{\{i,j\}}$), 
\item[$b)$] $z$ is adjacent to all $t_i$ (due to voter $v^{t_i}$),
\item[$c)$] $\{S_1,t_1\}$ is in the graph (due to $w$).
\end{itemize}
 We call these edges necessary edges. We claim that there exists a set cover of size (at most) $k$ iff there is a graph compatible with the profile with degree at most $\nbCandidates+k$.

Suppose that there is a set cover $\mathcal{C}$ of size at most $k$. Then, beyond the necessary edges mentioned above, we put an edge $\{S_i,z\}$ iff $S_i$ is in $\mathcal{C}$. Then the vertex with maximum degree is $z$, with degree $\nbCandidates+k$. The graph $\mathcal{G}$ is compatible with each voter $v_i$ because $\mathcal{C}$ is a set cover. It is compatible with voter $v_{\{i,j\}}$ thanks to the necessary edges (and $k\!>\!0$, so $z$ is connected as well). It is compatible with $v^{t_i}$ and $w$ thanks to the necessary edges.

Now suppose that there is a solution $\mathcal{G}$ with degree at most $\nbCandidates+k$. In particular, $z$ has degree at most $\nbCandidates+k$, hence is adjacent to at most $k$ vertices $S_i$. The preference of voter $v_i$ imposes that $z$ is adjacent to some $S_i$ which contains $e_i$. In other words, the set $\mathcal{C}$ of these (at most) $k$ sets $S_i$ is a set cover of size at most $k$. 
\end{proof}

\section{RECOGNITION OF TREES AND PATHS}\label{sec:lp}
In this section, we focus on the tree and path recognition - given a profile $\mathcal{P}$, we are looking for a tree (or a path) on which the profile if SP. 

Recognizing single-peaked preferences on a tree can be done using the procedure by Trick~\cite{trick1989recognizing}. As an alternative proof of this result, we show in this section that the continuous relaxation of the ILP formulations given in Section~\ref{sec:ILP} can be used to solve this recognition problem in polynomial time: in fact, all (optimal) extremal solution are integral (Theorem~\ref{th:tree}). We show in Theorem~\ref{th:path} a similar result for the recognition of profiles SP on a path. 

We start by recalling Trick's procedure~\cite{trick1989recognizing}, as we will use it in the proof of the results of the two theorems mentioned above. 

\paragraph{Recognition of profiles SP on a tree~\cite{trick1989recognizing}}


Let $\mathcal{P}$ be a profile containing preference lists $\lbrace R_1, \ldots , R_{\nbVoters} \rbrace $ of $\nbVoters$ voters over $\nbCandidates$ candidates. The algorithm of Trick builds a tree in an iterative way. In each iteration, it identifies one candidate (i.e., graph vertex) which is necessarily a leaf, and it determines the set of vertices this leaf can be connected to. It deletes then this leaf from all preferences, and repeats this process on the modified profile with preferences over $\nbCandidates\!-\!1$ candidates. The algorithm stops when only one candidate is remaining. Let us describe in more details the leaf recognition :

Let $k$ be a candidate placed at the last position by at least one voter. Trick shows that, if preferences are SP on a tree, then $k$ must necessarily be a leaf. More formally, for each $i\!\in\!\lbrace 1, \ldots , \nbVoters \rbrace$, let us denote by $A(k)_i$ the set of candidates ranked better than $k$ by voter $v_i$ if $k$ is not ranked first by $v_i$; if $k$ is ranked first by $v_i$, then $A(k)_i$ is the singleton containing the second most-preferred candidate of $v_i$. From $A(k)\!=\! \bigcap_{i = 1}^{\nbVoters} A(k)_i$, the following conclusions can be drawn:
\begin{itemize}
    \item if $A(k)\!=\! \emptyset$, there does not exist a tree solution.
    \item Otherwise, $A(k)$ is the set of vertices the leaf $k$ can be connected to.   
\end{itemize}

\begin{ex}
Consider the profile $\mathcal{P}\!=\!\{R_1,R_2,R_3\}$ defined by: 
\begin{align*}
R_1 : & \ (1, 2, 3, 4) \\
R_2 : & \ (2, 1, 3, 4) \\
R_3 : & \ (4, 1, 2, 3) \\
\end{align*}
\vspace{-1.25cm}
\begin{itemize}
    \item[1.] The candidate $4$ is classed at the last position by at least one voter - we will determine the set $A(4)$:\\ [0.5ex]
    $~~~~~~~~~~~~~~~~A(4) = \lbrace 1,2,3\rbrace \cap \lbrace 1,2,3\rbrace  \cap \lbrace 1\rbrace =  \lbrace 1 \rbrace $\\ [0.5ex]
    The candidate $4$ is then deleted from the preference lists and we continue next iteration with the sub-profile 
    $R_1^1= (1,2,3)$, $R_2^1= (2,1,3)$ and
    $R_3^1 =(1,2,3)$.
    
    \item[2.] The candidate $3$ is classed at the last position by at least one voter - we see that $A(3) = \lbrace 1,2 \rbrace $. We continue with the sub-profile  $R_1^2 = (1,2)$, $R_2^2 =(2,1)$, $R_3^2 =(1,2)$.
    \item[3.] We get $A(2) = \lbrace 1 \rbrace$, and the algorithm stops as we obtain a sub-profile involving only one candidate. 
\end{itemize}

To sum up, we have obtained $A(2) = \lbrace 1 \rbrace$ (first iteration), $A(3) = \lbrace 1, 2 \rbrace$ (second iteration) and $A(4) = \lbrace 1 \rbrace $ (third iteration). Consequently, in any tree compatible with $\mathcal{P}$, vertex 2 and vertex 4 have to be connected to vertex 1, and vertex 3 has to be connected to vertex~1 or 2. Hence, there exists two trees on which profile $\mathcal{P}$ is SP, and these are:
\begin{center}
\begin{tikzpicture}[scale = 0.6]
    \node[shape=circle,draw=black] (1) at (0,3) {1};
    \node[shape=circle,draw=black] (2) at (3,3) {2};
    \node[shape=circle,draw=black] (3) at (3,0) {3};
    \node[shape=circle,draw=black] (4) at (0,0) {4};

    \path [-] (1) edge (2);
    \path [-](1) edge (4);https://fr.overleaf.com/3458313817pdmxzrcqghdj
    \path [-] (3) edge (1);
    
    \node[shape=circle,draw=black] (1') at (6,3) {1};
    \node[shape=circle,draw=black] (2') at (9,3) {2};
    \node[shape=circle,draw=black] (4') at (6,0) {4};
    \node[shape=circle,draw=black] (3') at (9,0) {3};

    \path [-] (1') edge (2');
    \path [-](1') edge (4');
    \path [-] (3') edge (2');
    
\end{tikzpicture}
\end{center}
\end{ex}






\paragraph{Using LP to recognize SP preferences on a tree or a path}

Let us consider the following continuous relaxation LP-SP (linear program for single-peakedness) of the ILP introduced in Section~\ref{sec:ILP}:
\begin{align*}
 \min \ & \textstyle\sum_{\{k,l\}\subseteq\{1,\ldots,m\}} x_{\{k,l\}}\\
\mbox{s.t.} \ & \begin{cases}
\sum_{j = 1}^{k-1} x_{\{\pi_i(j), \pi_i(k)\}} \geq 1 \ \ \forall i\!\in\!\lbrace 1, \hdots, \nbVoters \rbrace, k \!\in\!  \lbrace 2, \hdots, \nbCandidates \rbrace \\
x_{\lbrace k, l \rbrace}\!\in\! [ 0, 1 ] \ \forall \lbrace k, l \rbrace\!\subseteq\!\lbrace 1, \hdots , \nbCandidates \rbrace
\end{cases}
\end{align*}

We show in Theorem~\ref{th:tree} that we can use LP-SP to solve in polynomial time the problem to determine, given a profile, whether there exists or not a tree compatible with it. 

\begin{thm}\label{th:tree}
If a profile $\mathcal{P}$ is compatible with a tree, then any extremal optimal solution $x$ of LP-SP is integral, i.e., $x_{\{k,l\}}\!\in\!\{0,1\}$ for any $\{k,l\}\!\subseteq\!\{1,\ldots,m\}$.
\end{thm}

\proof

The proof is based on two properties of optimal solutions of LP-SP when the profile is compatible with a tree. These two properties allow to come up with a reformulation of the problem as a maximum flow problem, where there is a bijection between the solutions of LP-SP of value $\nbCandidates-1$ and the (optimal) flows of value $\nbCandidates-1$. The result then comes from the fact that any extremal solution of the flow problem (with integral capacity) is integral \cite{ahuja1988network}.

The first property states that all constraints of LP-SP are tight in a solution of value $\nbCandidates-1$.

\medskip

\noindent\textbf{Property 1.} If the optimal value of LP-SP is $\nbCandidates\!-\!1$, then all constraints are tight in an optimal solution $x^*$: $\sum_{j = 1}^{k-1} x^*_{\{\pi_i(j), \pi_i(k)\}}\!=\!1$.

\medskip

\noindent {\it Proof of Property 1.} Let $v_i$ be a voter. There are $\nbCandidates\!-\!1$ constraints associated with $v_i$, and each variable $x_{\lbrace k,l \rbrace }$ appears  in exactly one of these constraints. Since on the one hand the sum of all variables is $\nbCandidates\!-\!1$ (objective function), and on the other hand the sum of variables in each of these $\nbCandidates\!-\!1$ constraints is at least one, each constraint must be tight. This concludes the proof of Property 1.\\  

Now, let us consider that the profile is SP with respect to a tree. The recognition procedure recalled above starts by identifying a candidate, say $\nbCandidates$, ranked last in at least one ranking and such that $A(\nbCandidates)\!\neq\!\emptyset$. This procedure is then applied recursively, till there is only one candidate. For simplicity, let us assume that the first removed (identified) candidate is $\nbCandidates$, the second $\nbCandidates\!-\!1$, and so on. To avoid confusion, we denote $B(k)$ the set $A(k)$ when considering the profile restricted to the first $k$ candidates (when $k$ is identified as a leaf, and then removed from the profile). 

\medskip

\noindent\textbf{Property 2.} If the profile is SP on a tree, then in an optimal solution of LP-SP, for any candidate $k\!\geq\!2$ we have $\sum_{j \in B(k) } x_{\{j,k\}} \!=\!1$, and $x_{\{j,k\}}\!=\!0$ for any $j\!\in\!\{1,\dots,k-1\}\setminus B(k)$.

\medskip

\noindent {\it Proof of Property 2.} 
Let us consider some candidate $k\geq 2$, some optimal solution $X$ of LP-SP, and assume that the properties in the lemma are true for any $k'>k$. We will show that they are also true for $k$. 

To do this, let us define LP-SP($k$) as the linear program corresponding to the problem restricted to the candidates $1,2,\dots,k$. We first need to show that the optimal solution $X$ restricted to the first $k$ candidates, let us call it $X_k$, is feasible and optimal for LP-SP($k$). 
To do this, let us consider a constraint of LP-SP($k$), let us say the constraint of connecting  candidate $j\leq k$ for some voter $i$. In the corresponding constraint in the initial program LP-SP, there are possibly some other variables $x_{k',j}$: there is such variable $x_{k',j}$ for each candidate $k'>k$ ranked before $j$ by $i$. But then when $k'$ was removed, if $x_{k',j}$ appears in the constraint then $j\not \in B(k')$, and then $x_{k',j}=0$. So all `removed variables' in the constraint was set to 0 in $X$, hence $X_k$ is feasible for LP-SP($k$). We can now easily see that it is optimal: each time a candidate $k'>k$ has been removed, $\sum_{j\in B(k')}x_{j,k}=1$ so the total weights of (remaining) variables reduce by 1. Thus $X_k$ is a feasible solution of LP-SP($k$) of value $k-1$.

Now we can focus on $X_k$ on LP-SP($k$). Note that the profile is trivially SP on the first $k$ candidates (as it is SP on the whole set of candidates). Candidate $k$ is ranked in last position by some voter $v_i$, so we have $\sum_{j=1}^{k-1}x_{\{j,k\}}\!\geq\!1$ (constraint of connecting $k$ for voter $v_i$), and by Property 1 we have $\sum_{j=1}^{k-1}x_{\{j,k\}}\!=\!1$. If all  candidates $1,2,\dots,k-1$ are in $B(k)$ then we are done. Otherwise, consider a candidate $\ell\!\not\in\!B(k)$. Then $\ell$ is ranked after $k$ by some other voter $v_j$ (and, for this voter $v_j$, $k$ and $\ell$ are not the best two candidates). Then, if  $x_{\{\ell,k\}}\!>\!0$ we get $\sum_{j\not\in \{\ell,k\}} x_{\{j,k\}}\!=\!1\!-\!x_{\{\ell,k\}}\!<\!1$, and the constraint associated to $v_j$ for connecting $k$ to its predecessors is violated. So $x_{\{\ell,k\}}\!=\!0$ for any $\ell\!\not\in\!B(k)$, and consequently  $\sum_{j \in B(k) } x_{\{j, k\}}\!=\!1$.
This concludes the proof of Property 2.\\

Now we reformulate the problem as a flow problem. From $\mathcal{P}$, we build a network (directed graph) $R$ with:
\begin{itemize}
    \item A source $s$, a destination $t$, and for each candidate $k$ two vertices $\ell_k$ and $r_k$.
    \item We have an arc from $s$ to each $\ell_k$ with capacity 1, and an arc from each $r_k$ to $t$ with capacity $\infty$.
    \item For each candidate $k$, we have an arc $(\ell_k,r_j)$ for each $j\!<\!k$. The capacity of this arc is 1 if $j\!\in\!B(k)$, and 0 otherwise.
\end{itemize}
Let us denote by $\phi$ a flow on this network, with $\phi(e)$ the flow on edge $e$. Note that $\ell_1$ has no outgoing edge, so the optimal flow is at most $\nbCandidates\!-\!1$.

We show that the correspondence $x_{\{k,j\}}\!=\!\phi(k,j)$ (for each $j\!<\!k$) is a bijection between solutions of value $\nbCandidates-1$ of LP-SP and (optimal) flows of value $\nbCandidates\!-\!1$ in $R$.

Let $\phi$ be a flow of value $\nbCandidates-1$. As there is no flow through $\ell_1$, there is a flow of value 1 through each $\ell_k,k\!>\!1$. Since arc $(k,j)$ has capacity 0 if $j\!\not\in\!B(k)$, by flow conservation we have $\sum_{j\in B(k)}\phi(k,j)\!=\!1$, which means that $\sum_{j\in B(k)}x_{\{k,j\}}\!=\!1$. Now consider a voter $v_i$ where $k$ is not ranked first. By the procedure of Trick, when $k$ is identified as a leaf, all candidates in $B(k)$ are ranked before $k$, and the corresponding constraint is satisfied. This is true for all candidates and voters, so $x$ is a feasible solution of LP-SP, of value $\nbCandidates\!-\!1$.

Conversely, let $x$ be a feasible solution of LP-SP of value $\nbCandidates\!-\!1$. From Property 2, we have $\sum_{j \in B(k) } x_{\{j,k\}}\!=\!1$ for each candidate $k\geq 2$. This immediately gives a flow of value $\nbCandidates\!-\!1$. 

By integrality of extremal flows (any non integral optimal flow is a convex combination of integral flows), any extremal optimal solution of LP-SP is integral (when there exists a tree compatible with $\mathcal{P}$).
\endproof

Let us now turn to the recognition of profiles SP on a path. A (connected) graph is a path iff it is a tree with degree at most~2. Hence, we consider the following ILP formulation where we minimize the number of edges and add constraints on the vertex degrees:
\begin{align*}
 \min \ & \sum_{\lbrace k, l \rbrace \subset \lbrace 1, \ldots, \nbCandidates \rbrace} x_{\lbrace k, l \rbrace} \\
\mbox{s.t.} \ & \begin{cases}
\sum_{j = 1}^{k-1} x_{\{\pi_i(j), \pi_i(k)\}} \geq 1 \ \ \forall i\!\in\!\lbrace 1, \hdots, \nbVoters \rbrace, k\!\in\!\lbrace 2, \hdots, \nbCandidates \rbrace \\ \\
\sum_{l =1, l\neq k}^{\nbCandidates} x_{\lbrace k, l \rbrace } \leq 2 \ \  \forall k\!\in\!\lbrace 1, \hdots , \nbCandidates \rbrace \\ \\
x_{\lbrace k, l \rbrace}\!\in\!\lbrace 0, 1 \rbrace \ \forall \lbrace k, l \rbrace\!\subseteq\!\lbrace 1, \hdots , \nbCandidates \rbrace
\end{cases}
\end{align*}
Clearly, a profile is compatible with a path iff the optimal value of the previous ILP is $\nbCandidates\!-\!1$. Let us call LP-SP2 the continuous relaxation. By using very similar arguments as above (same reformulation as a flow problem), one can prove the following result. 


\begin{thm}\label{th:path}
If a profile $\mathcal{P}$ is compatible with a path, then any extremal optimal solution of LP-SP2 is integral, i.e., $x_{\{k,l\}}\in\{0,1\}$ for any $\{k,l\}\!\subseteq\!\{1,\ldots,m\}$.
\end{thm}

\section{RECOGNITION OF PSEUDOTREES}
\label{sec:Pseudotrees}
So far, we have seen that our minimization problem is NP-hard in the general case, but polynomially solvable in the case where the optimal solution is a tree. As a natural extension, we consider the problem to recognize profiles that are single-peaked with respect to a graph with $\nbCandidates-1+k$ edges, for some fixed $k$, thus allowing $k$ more edges than in a tree. In this section, we consider the case $k=1$. A graph on $\nbCandidates$ vertices with $\nbCandidates$ edges is called a pseudotree. We show that recognizing if there exists a pseudotree compatible with a given profile can be done in polynomial time. We leave as open question the parameterized complexity of the problem when $k$ is the parameter: would the problem be in XP? Or even in FPT?


Let us now deal with  the case of pseudotree. Hence, the set of solutions we want to recognize is the class of connected graphs having (at most) $\nbCandidates$ edges. 
To solve the problem in polynomial time, we devise an algorithm that first identifies the leaves of the pseudotree and then the cycle on the remaining vertices. The second step (cycle recognition) is done using the polynomiality of recognizing single-peakedness on a cycle \cite{peters17preferences}. For the first step, we need to modify the procedure of recalled in Section~\ref{sec:lp}. This procedure was able to correctly identify leaves when the profile was compatible with a tree, but it fails to correctly identify leaves when the underlying structure is a pseudotree. With a slight modification though, we obtain in Proposition~\ref{prop:pt1} a necessary and sufficient condition for a candidate to be a leaf in a pseudotree. This is the stepping stone leading to the polynomiality of detecting whether a given profile is compatible with a pseudotree, stated in Theorem~\ref{th:pt}.

\begin{ex}\label{ex:pt}
Let us consider the profile on 4 voters and 5 candidates given in Example~\ref{ex21}, for which there is a (unique) pseudotree compatible with it. 

The procedure to detect leaves when looking for a tree focuses on candidates ranked last in some $R_i$, candidates $1$ and $5$ here, and $A(1)=A(5)=\emptyset$. Note that the whole profile is not compatible with a cycle, so we need somehow to first detect $4$ as a leaf, and then detect that the candidates $1,2,3,5$ are SP with respect to a cycle.
\end{ex}
The central property that allows to recognize profiles compatible with a pseudotree is given in the following proposition.
\begin{prop}\label{prop:pt1}
Let $\mathcal{P}$ be a preference profile, and suppose that a candidate $i$ is such that $A(i)\neq \emptyset$. Then $\mathcal{P}$ is compatible with a pseudotree if and only if it is compatible with a pseudotree where $i$ is a leaf. 
\end{prop}

\proof

Let $\mathcal{G}$ be a pseudotree compatible with $\mathcal{P}$   where $i$ is not a leaf. We transform $\mathcal{G}$ into a pseudo-tree $\mathcal{G}'$ compatible with $\mathcal{P}$ where $i$ is a leaf. Let $j\in A(i)$.

{\bf Case 1: $\{i,j\}\in \mathcal{G}$.} Let us first consider an easy case, where $\{i,j\}\in \mathcal{G}$. Then we build $\mathcal{G}'$ from $\mathcal{G}$ by simply replacing each edge $\{i,k\}$ (with $k\neq j$) by the edge $\{j,k\}$. Since for each voter either $j$ is ranked before $i$, or $i$ is first and $j$ second, then this modification creates a graph $\mathcal{G}'$ compatible with all the preferences. Note that $\mathcal{G}'$ has (at most) as many edges as $\mathcal{G}$, so it is a pseudotree (or a tree, and we can add any edge to create a pseudotree).

{\bf Case 2: $\{i,j\}\not\in \mathcal{G}$.} 
Let us now consider the case where $\{i,j\}\not \in \mathcal{G}$. Note that then $j$ is ranked before $i$ in all preferences (otherwise $i$ is first and $j$ is second, and the edge $\{i,j\}$ is forced to be in any compatible graph, a contradiction). Then we transform $\mathcal{G}$ into a graph $\mathcal{G}'$ which is a pseudotree containing the edge $\{i,j\}$, and then Case 1 applies to $\mathcal{G}'$. To do this, let us consider two subcases.

{\bf Case 2a.} If, in $\mathcal{G}$, in all (simple) paths from $j$ to $i$ the predecessor of $i$ is the same vertex $u$. Then we create $\mathcal{G}'$ by replacing the edge $\{u,i\}$ by the edge $\{j,i\}$. Consider a voter $v$. Since $j$ is ranked before $i$ by $v$, then $u$ is ranked before $i$ by $v$ (the subgraph induced by $i$ and the candidates ranked before him by $v$ is connected and contains $i$ and $j$, so it contains $u$). Then the modification does not affect $u$ (it is still connected to one of the candidates ranked before him), and $i$ is now connected to $j$.

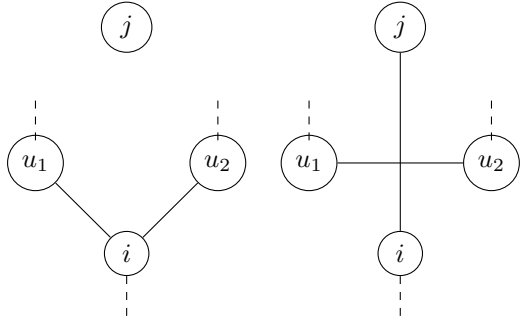
\begin{figure}[ht]\label{fig:edgeswap2}
\begin{center}
\begin{tikzpicture}[scale=0.6]
    \node[shape=circle,draw=black] (1) at (2,5) {$j$};
    \node[shape=circle,draw=black] (2) at (0,2) {$u_1$};
    \node[shape=circle,draw=black] (3) at (4,2) {$u_2$};
    \node[shape=circle,draw=black] (4) at (2,0) {$i$};
 
    \node[shape=circle,draw=black] (5) at (8,5) {$j$};
    \node[shape=circle,draw=black] (6) at (6,2) {$u_1$};
    \node[shape=circle,draw=black] (7) at (10,2) {$u_2$};
    \node[shape=circle,draw=black] (8) at (8,0) {$i$};

    \path [-] (2) edge (4);
    \path [-] (3) edge (4);
    \path [-] (6) edge (7);
    \path [-] (5) edge (8);
    \draw [dashed] (0,2.5) -- (0,3.5);
    \draw [dashed] (4,2.5) -- (4,3.5);
    \draw [dashed] (6,2.5) -- (6,3.5);
    \draw [dashed] (10,2.5) -- (10,3.5);
    \draw [dashed] (2,-0.5) -- (2,-1.5);
    \draw [dashed] (8,-0.5) -- (8,-1.5);
            

\end{tikzpicture}
\end{center}
\caption{Edge swap}
\end{figure}

{\bf Case 2b.} In the other case, in $\mathcal{G}$ there are two simple paths from $j$ to $i$ such that the predecessor of $i$ is $u_1$ in the first one and $u_2\neq u_1$ in the second one (note that there cannot be more than 2 since $\mathcal{G}$ is a pseudotree). We build $\mathcal{G}'$ from $\mathcal{G}$ by deleting the edges $\{u_1,i\}$ and $\{u_2,i\}$, and adding edges $\{i,j\}$ and $\{u_1,u_2\}$, see Figure~\ref{fig:edgeswap2}.

Consider a voter $v$. Since $v$ prefers $j$ to $i$ and the subgraph of $\mathcal{G}$ induced by the candidates up to $i$ in the ranking of $v$ is connected, then $u_1$ or $u_2$ is ranked before $i$ by $v$, say $u_1$ (we assume wlog that $u_1$ is preferred to $u_2$ by $v$). Then we see that $\mathcal{G}'$ is compatible with the preference of $v$: indeed, when considering candidates one by one in the order of $v$, the only modification holds for $u_2$, which is now connected to $u_1$ (ranked before him), and for $i$, which is now connected to $j$ (ranked before him).
\endproof
Note that $i$ can be connected to any vertex $j\in A(i)$.

Before giving the procedure that recognizes preferences compatible with a pseudotree, we need to establish another property regarding such preferences.
\begin{prop}\label{prop:pt2}
 If  a preference profile $\mathcal{P}$ is compatible with a pseudotree, then either there exists a candidate $i$ such that $A(i)\neq \emptyset$, or $\mathcal{P}$ is compatible with a cycle.
\end{prop}
If $\mathcal{P}$ is not compatible with a cycle, then there exists a leaf $i$ in a pseudotree $\mathcal{G}$ compatible with $\mathcal{P}$. Let $j$ be the unique neighbor of $i$ in $\mathcal{G}$. Suppose that $j\not\in A(i)$. Let $v$ be a voter who prefers $i$ to $j$, and $j$ is not second if $i$ is first in the ranking of $v$. If $i$ is not first, then the subgraph induced by the candidates up to $i$ (in the ranking of $v$) is not connected. If $i$ is first (and $j$ not second), the subgraph induced by the first two candidates (in the ranking of $v$) is not connected. Contradiction.
\endproof

Consider now the following procedure \textsc{Detect\_PseudoTree} that detects pseudotree-singlepeakedness:
\begin{enumerate}
    \item Set $E'=\emptyset$
    \item While there are at least 4 candidates, and a candidate $i$ such that $A(i)\neq \emptyset$:
    \begin{enumerate}
        \item Add edge $\{i,j\}$ to $E'$ for some $j\in A(i)$. 
        \item Remove $i$ from the profile.
    \end{enumerate}
    \item Detect if there is a cycle $C$ which is compatible with the (remaining) profile:
    \begin{itemize}
        \item If YES: output $E'\cup C$
        \item If NO: output NO.
    \end{itemize}
\end{enumerate}

\begin{thm}\label{th:pt}
Given a preference profile $\mathcal{P}$ on at least 3 candidates, \textsc{Detect\_PseudoTree} is a polytime procedure which outputs a pseudotree compatible with $\mathcal{P}$ if some exists, and outputs NO otherwise.
\end{thm}

\proof
\textsc{Detect\_PseudoTree} obviously runs in polynomial time. We proceed by induction on the number of candidates. If there are three candidates the procedure outputs a cycle on these 3 candidates. Now suppose that the result is true up to $\nbCandidates-1$ candidates, and consider a profile $\mathcal{P}$ on $\nbCandidates\geq 4$ candidates. 

Suppose that $\mathcal{P}$ is compatible with a pseudotree $\mathcal{G}$. 
\begin{itemize}
    \item If there exists a candidate $i$ with $A(i)=\emptyset$, then by Proposition~\ref{prop:pt1}, there exists a pseudotree $\mathcal{G}'$ compatible with $\mathcal{P}$ where $i$ is a leaf. Then the profile obtained from $\mathcal{P}$ by removing $i$ is compatible with a pseudotree ($\mathcal{G}'\setminus{i}$), and adding the edge $\{j,i\}$ as done by \textsc{Detect\_PseudoTree} gives a pseudotree compatible with $\mathcal{P}$.
    \item Otherwise, by Proposition~\ref{prop:pt2}, $\mathcal{P}$ is compatible with a cycle, which is found by \textsc{Detect\_PseudoTree} (Step 3).
\end{itemize}

Suppose now that \textsc{Detect\_PseudoTree} does not output NO. If there were no candidate $i$ with $A(i)\neq \emptyset$, then $\mathcal{P}$ is compatible with the cycle $C$. Otherwise, let $i$ be the candidate in the first iteration of the loop in Step 2 ($A(i)\neq \emptyset$). 
Then, on the profile $\mathcal{P}$ where $i$ is removed, \textsc{Detect\_PseudoTree} outputs a pseudotree, compatible with this profile without $i$ by induction. Since $j\in A(i)$, adding edge $\{j,i\}$ makes the pseudotree compatible with $\mathcal{P}$.
\endproof

Note that the generalization of this polynomiality result to, say, connected graphs with $(n\!-\!1\!+\!k)$ edges seems to require new techniques (even for fixed $k$, i.e. to show that the problem is in XP when parameterized by $k$). Indeed, an enumeration of all subsets of $k$ edges does not allow to reduce the problem to trees. Procedure \textsc{Detect\_PseudoTree} does not seem to generalize as well, as it specifically relies on the decomposition of the solution into one cycle and leaves.

\section{EXPERIMENTAL STUDY}
\label{sec:ExperimentalStudy}
We carried out numerical experiments\footnote{All tests were performed on a Intel Core i7-1065G7 CPU with 8 GB of RAM under the Windows operating system. We used the IBM Cplex solver for the solution of ILPs.} on real and randomly generated instances of the problems tackled in the paper. In the case of real data, we compare the optimal solution of the ILP to that of its continuous relaxation. We also focus on the ability to detect structure in voters' preferences depending on the election context. To go further, we use randomly generated instances to study structural aspects of solutions; we notably study the graph density depending on the number of voters and on the dispersion of their opinions.

\subsection{Numerical tests on real data}
We used PrefLib data sets available on \texttt{www.preflib.org} to perform our numerical tests~\cite{MaWa13a}. While this database offers four different types of data, only the ED (Election Data) type is relevant for our study. Among the ED data sets, we used the complete strict order lists (which correspond to files with \texttt{.soc} extension). 

At the time we carried out these experiments, 315 data files of this type were available in PrefLib, however, many of them were not adapted to our study for several reasons. The first one is that many elections dealt with only 3 or 4 candidates and a great number of voters, hence the obtained graph was, unsurprisingly, always complete. 
We also met the opposite problem when there were very few voters, typically 4, so there was no point in looking for some general structure. 
Thus, in practice, there were 25 real data files usable for our purposes, namely:
\begin{itemize}
    \item 20 files from the \texttt{ED-00006} data set, which contains figure skating rankings from various competitions during the 1998 season including the World Juniors, World Championships, and the Olympics.
    \item 2 files from the \texttt{ED-00009} data set, which contains the results of surveying students at AGH University of Science and Technology (Krakow, Poland) about their course preferences.
    \item 1 file from the  \texttt{ED-00012} data set, where individuals ranked T-shirt designs. 
    \item 1 file from the \texttt{ED-00014} data set, which contains preferences about various kinds of sushi (surveys conducted by Toshihiro Kamishima).
    \item 1 file from the \texttt{ED-00032} data set, which contains the results of surveying students in the Faculty of Informatics, Instituto Superior Politécnico José Antonio Echeverría (Cujae, Havana, Cuba), about the most important criteria affecting their performances as students.
\end{itemize}

\noindent We now present the results obtained for these 25 instances. The tackled optimization problem was to determine a graph with a \emph{minimal number of edges}. 
For all instances considered here, an optimal graph has been obtained in about 40 milliseconds for the ILP formulation and 20 milliseconds for its LP relaxation. In fact, the linear programming formulation always returned an integer solution. 
Table~\ref{fig:tests_real} summarizes the obtained results.

\paragraph{\texttt{ED-00006} data set.} The number of candidates (skaters) varies from 14 to 30, and the number of voters (judges) from 7 to 9. For the 20 instances considered, a tree has been obtained 14 times, a pseudotree 5 times, and a solution with 23 edges  for 19 candidates (vertices) has been obtained once. The possible interpretation of these results is that, even though the rankings are based on subjective opinions of the judges, there is something like a ``true ranking'' behind as some skaters are objectively better than other ones. Thus, the rankings given by the judges can be viewed as biased observations of the true ranking, so that they are quite close. We note also that even when the solution was not a tree, the LP continuous relaxation gave an integer solution (identical to the one of the ILP). We also precise that we also checked compatibility with an axis, and no profile was single-peaked with respect to an axis. 

\paragraph{\texttt{ED-00009} data set.} Each student provided a rank ordering over all the courses with no missing elements. There were 9 courses to choose from in 2003 and 7 in 2004, and about 150 students. For both years, the optimal solution was a tree, more specifically a star. This is easily explained from the fact that, in both years, there was one course which was the most preferred for every student.  

\paragraph{\texttt{ED-00012} and \texttt{ED-000014} data sets.} The optimal solution for the preferences over the T-shirt designs had 25 edges, which is quite a lot regarding the number of candidates (11) and voters (30). However, it is consistent with the intuition that there is probably no structure behind T-shirt designs. The same can be said for the preferences over the kinds of sushi, where 5000 voters were asked for their preferences about 10 kinds of sushi (the optimal solution is a complete graph in this case). 

\paragraph{\texttt{ED-00032} data set} In the single instance with no tie nor missing element, there were 15 students that ranked the 6 criteria affecting their performances. In both the ILP and LP formulations, a solution with 8 edges has been returned.  

\begin{figure}[H]
    \centering
    \begin{center}
 \begin{tabular}{||c | c | c | c|} 
 \hline
 File &  \#candidates & \#voters & \#edges \\ [0.5ex] 
 \hline\hline
 \texttt{ED-00006-00000003.soc} &  14 & 9 & 13 (tree) \\ 
 \hline
 \texttt{ED-00006-00000004.soc} &  14 & 9 & 13 (tree) \\ 
  \hline
\texttt{ED-00006-00000007.soc} &  23 & 9 & 22 (tree) \\ 
 \hline
 \texttt{ED-00006-00000008.soc} &  23 & 9 & 22 (tree) \\ 
 \hline
 \texttt{ED-00006-00000011.soc} &  20 & 9 & 20 (pseudotree) \\ 
 \hline
 \texttt{ED-00006-00000012.soc} &  20 & 9 & 20 (pseudotree) \\ 
 \hline
 \texttt{ED-00006-00000018.soc} &  24 & 9 & 23 (tree) \\ 
 \hline
 \texttt{ED-00006-00000021.soc} &  18 & 7 & 17 (tree) \\ 
 \hline
 \texttt{ED-00006-00000022.soc} &  18 & 7 & 17 (tree) \\ 
 \hline
 \texttt{ED-00006-00000028.soc} &  24 & 9 & 23 (tree) \\ 
 \hline
  \texttt{ED-00006-00000029.soc} &  19 & 9 & 23 \\
 \hline
  \texttt{ED-00006-00000032.soc} &  23 & 9 & 23 (pseudotree) \\ 
 \hline
  \texttt{ED-00006-00000033.soc} &  23 & 9 & 22 (tree) \\ 
 \hline
   \texttt{ED-00006-00000034.soc} &  23 & 9 & 22 (tree) \\ 
 \hline
   \texttt{ED-00006-00000035.soc} &  18 & 9 & 17 (tree) \\ 
 \hline
  \texttt{ED-00006-00000036.soc} &  18 & 9 & 17 (tree) \\ 
 \hline
  \texttt{ED-00006-00000037.soc} &  19 & 9 & 18 (tree) \\ 
 \hline
 \texttt{ED-00006-00000044.soc} &  20 & 9 & 19 (tree) \\ 
 \hline
 \texttt{ED-00006-00000046.soc} &  30 & 9 & 30 (pseudotree) \\ 
 \hline
 \texttt{ED-00006-00000048.soc} &  24 & 9 & 23 (tree) \\ 
 \hline
 \texttt{ED-00009-00000001.soc} &  9 & 146 & 8 (tree) \\ 
 \hline
 \texttt{ED-00009-00000002.soc} &  7 & 153 & 6 (tree) \\ 
 \hline
 \texttt{ED-00012-00000001.soc} &  11 & 30 & 25 \\ 
 \hline
 \texttt{ED-00014-00000001.soc} &  10 & 5000 & 45 (clique)  \\
 \hline
 \texttt{ED-00032-00000002.soc} &  6 & 15 & 7 \\ 
 \hline
\end{tabular}
\end{center}
    \caption{Minimal number of edges on real data sets from PrefLib. The first column gives the file name, the second and third the number of candidates and voters. The fourth column gives the minimal number of edges; specific structures (axis, tree, pseudotree or clique) are indicated in parentheses when they appear.}
    \label{fig:tests_real}
\end{figure}

\subsection{Experimental study on randomly generated data}
The experimental study on real data revealed some interesting information. Nevertheless, it is limited by the small amount of data available. It seems indeed hard to obtain real election data with complete strict order rankings. To overcome this difficulty, other experiments can be considered. One of them is to adapt the approach to partial order rankings that can be met more often in practice. However, in this paper, we preferred to generate random data in order to study the structure of solutions and the relation between the solutions returned by ILP and LP formulations. 

As mentioned in the previous section, it seems that in some contexts we can assume that the voter's preferences are biased observations of a ``true'' ranking. This idea can be modeled using the Mallows distribution on rankings. In this model, the ``true'' ranking is called \emph{central permutation} and its probability is the highest one. The probability of other permutations decreases with the Kendall tau distance from the central permutation. Formally, let $R_0$ be the central permutation. The probability of a permutation $R$ is: 
$$ P(R) = \frac{\exp(-\theta d(R, R_0))}{\psi(\theta)}$$
where $d(.,.)$ is the Kendall tau distance, $\theta$  is a dispersion parameter modeling the opinion heterogeneity, and $\psi(\theta)$ is a normalisation constant. The parameter $\theta$ is a real number greater than or equal to 0. If $\theta = 0$, the uniform distribution is obtained. The greater the value of $\theta$, the more the voters agree on the central permutation. 

We used the PerMallows R package\footnote{\url{https://cran.r-project.org/web/packages/PerMallows/index.html}} for generating the random data according to the Mallows model. The number of candidates was set to $m\!=\!20$, the value of $\theta$ varied from 0 to 1 by step of 0.1. The number of voters $n$ varied from 20 to 100 by step of 10.   
For each pair $(\theta_0, n_0)$ of parameter values, the results are averaged over 1000 randomly drawn preference profiles. The curves in Figure~\ref{fig:density} shows the evolution of the graph density according to these parameter values.

\begin{table*}[p]
\begin{tabular}{c}
\begin{minipage}{\textwidth}
\begin{figure}[H]
    \centering
    \includegraphics[scale=0.35]{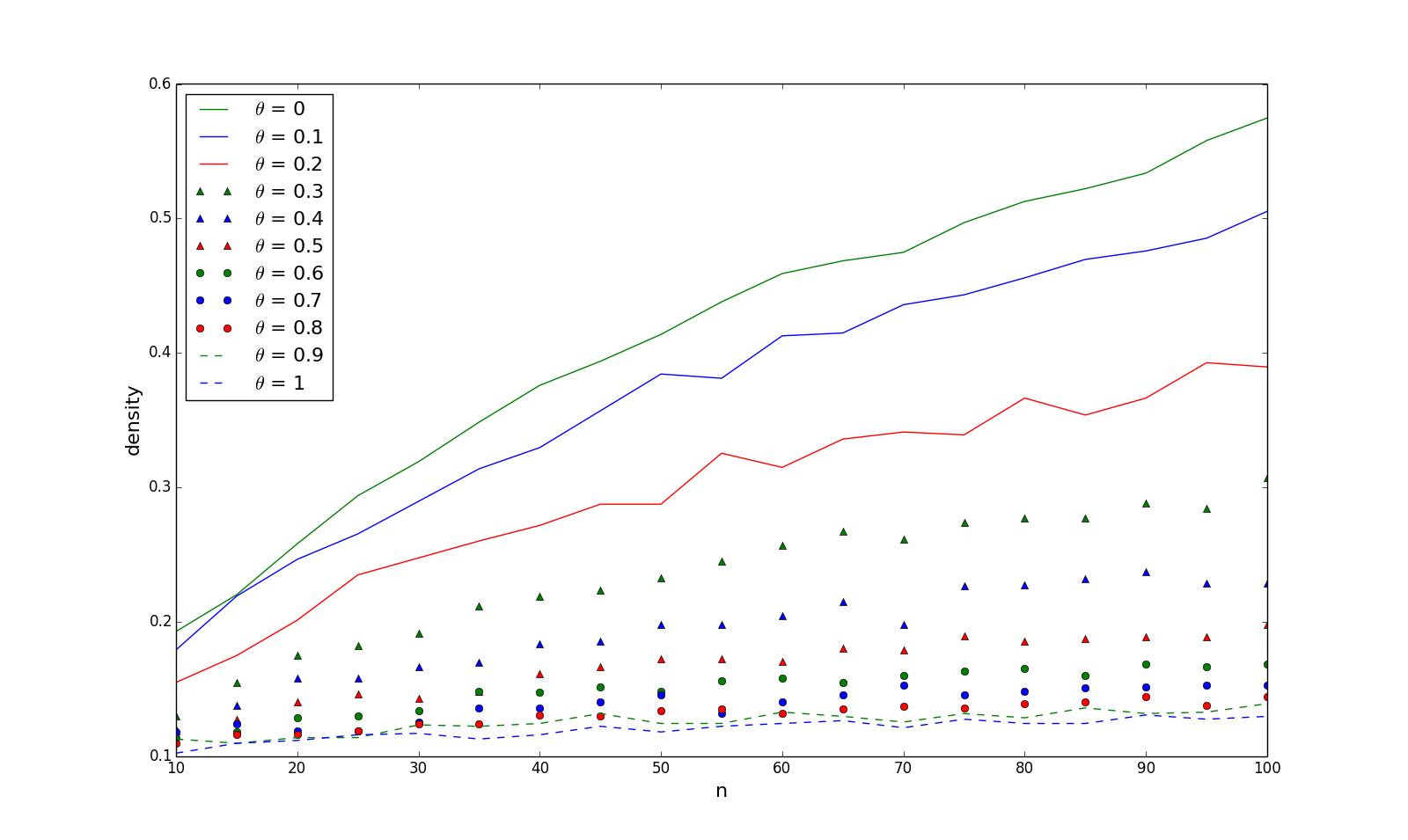}
    \caption{Density of the graph according to parameters $\theta$ and $n$ (with $m\!=\!20$). 
    }
    \label{fig:density}
\end{figure}
\end{minipage} \\
\begin{minipage}{\textwidth}
\begin{figure}[H]
    \centering
    \includegraphics[scale=0.35]{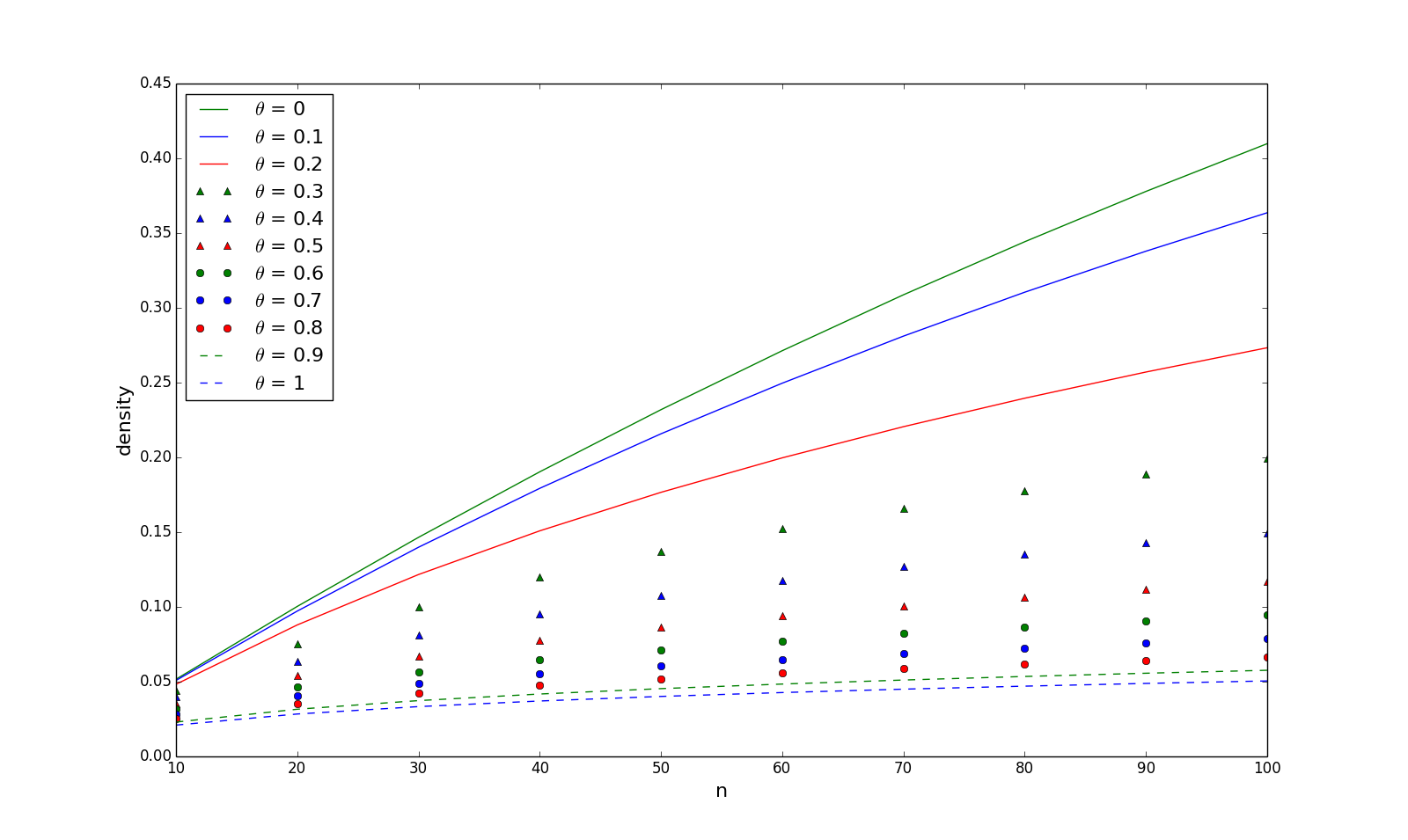}
    \caption{Expected number of necessary edges according to parameters $\theta$ and $n$ (with $m\!=\!20$).}
    \label{fig:density_proba}
\end{figure}
\end{minipage}
\end{tabular}
\end{table*}

In the best case, the obtained solution is a tree, hence, the density is $(m-1)/\frac{m (m-1)}{2}\!=\! \frac{2}{m}$. As we set $m = 20$, this corresponds to a density of 0.1. The function representing the graph density seems indeed to converge to the constant function of value 0.1 while the value of $\theta$ increases and the preferences in the profile become similar (the curves get closer and closer to the x-axis). Put another way, the density captures the similarity of voters' preferences, as clearly the higher $\theta$ the lower the curve. On the contrary, the graph density becomes of course higher when the number $n$ of voters increases. Nevertheless, note that, even for 100 voters, the graph is still quite far from being complete. Besides, the slope of the curve decreases with $n$. During our experiments, we plotted functions $1\!-\log(density)$ and obtained a set of (approximative) straight lines, thus indicating that the  convergence towards density 1 (complete graphs) is of the form $1\!-e^{-\lambda_\theta n}$, where $\lambda_\theta\!>\!0$ is a parameter decreasing with $\theta$.

We now give some theoretical arguments that support this observation. Let us recall that if a voter ranks $j$ first and $k$ second (or the opposite), then edge $\{j,k\}$ must be present in the graph and is called \emph{necessary} edge. Assuming that the preferences in the profile are generated with the Mallows model, let us now estimate the number of necessary edges in the graph for $n$ voters and $m$ candidates, which gives us an underestimation and hopefully good approximation of the total number of edges. Let $\theta$ be the model parameter and $R_0$ the central permutation. The probability that a preference induces the necessary edge $\{ j, k\}$ is 
\begin{equation} \label{eq:Pjk} 
P(\{j, k\}) = \frac{1}{\psi(\theta)}\sum_{R\in\mathcal{R}_{\{j,k\}}} \exp{(-\theta d(R, R_0))}
\end{equation}
where $\mathcal{R}_{\{j,k\}}$ is the set of permutations of $\{1,\ldots,m\}$ that ranks $j$ and $k$ in the first two positions.
In a profile with $n$ voters, the probability that no preference induces the necessary edge $\{j,k\}$ is then written
$$ {(1 - P(\{j,k\}))}^n.$$
Hence, by passing to the complement, the probability that $\{j,k\}$ is a necessary edge is 
$$ P_n(\{j,k\}) = 1 - {(1 - P(\{ j,k\}))}^n.$$
Finally, we obtain the expected value of the number of edges as 
\begin{equation} \label{eq:necessary_edges}
\sum_{\{j,k\} \subseteq \{ 1, \hdots , m \}} P_n(\{j,k\}) = \binom{m}{2} - \sum_{\{j,k\} \subseteq \{ 1, \ldots , m \}} \left(1 - P(\{j,k\})\right)^n.    
\end{equation}



For $\theta\!=\!0$, as the distribution is uniform, we get that $P(\{j,k\}) =1/{m \choose 2}$. Then, we directly obtain that the average number of necessary edges is ${m \choose 2}\left( 1-e^{-\alpha n}\right)$ with $\alpha\!=\!-\ln{\left(1-1/\binom{m}{2}\right)}$, thus contributing for $\left(1-e^{-\alpha n}\right)$ in the density, in accordance with the experiments. The curves in Figure \ref{fig:density_proba} shows the evolution of the expected contribution of necessary edges in the graph density according to the values of $\theta$ and $n$. The result extends to any value of $\theta$ but requires a dedicated algorithm to compute efficiently $P(\{j,k\})$ in Equation~\ref{eq:necessary_edges} (see Appendix~\ref{sec:necessary_edges}). 
As expected,
we can see that the shapes of the curves coincide in Figures~\ref{fig:density} and~\ref{fig:density_proba}. Note, however, that the scale of the y-axis in Figure~\ref{fig:density_proba} slightly differs from the one in Figure~\ref{fig:density} (the curves in Figure~\ref{fig:density_proba} indeed only account for \emph{necessary} edges, thus the analytical values are smaller than the experimental ones).

\begin{figure}[H]
    \centering
    \includegraphics[scale=0.35]{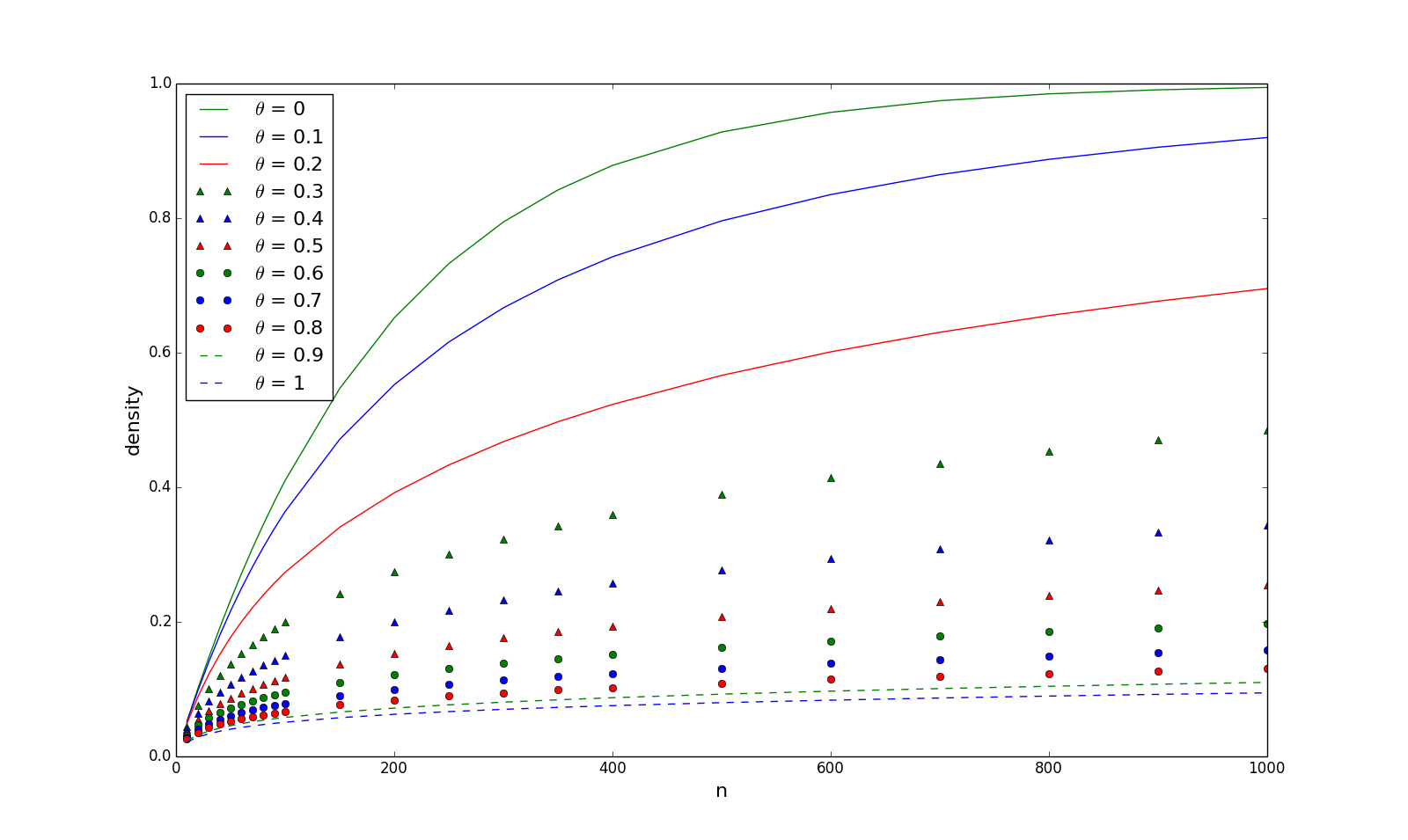}
    \caption{Expected number of necessary edges for up to $n\!=\!1000$ voters following the Mallows model of parameter $\theta$ (with $m=20$).}
    \label{fig:density_proba_1000}
\end{figure}

By using the formula in Equation~\ref{eq:necessary_edges}, we can have an idea of the evolution of the number of necessary edges in the graph for up to 1000 voters whose preferences follows the Mallows model. The obtained curves for various values of $\theta$ are shown in Figure~\ref{fig:density_proba_1000}. For instance, if one assumes that all votes are equally likely (\emph{impartial culture assumption}, corresponding to $\theta\!=\!0$), then the graph becomes complete for a thousand voters, while only around 45\% of the edges are present if one sets $\theta\!=\!0.2$ (i.e., a lower preference heterogeneity).


\section{CONCLUSION}

In this paper, we have investigated single-peakedness on an arbitrary graph $\mathcal{G}$, a generalization of single-peakedness of an axis that offers the advantage that any preference profile is at worst single-peaked on a complete graph. The recognition problem is then turned into an optimization problem the objective function of which depends on the type of structure that is looked for. After establishing negative complexity results regarding the minimization of the number of edges or the maximum vertex degree in $\mathcal{G}$, we have derived positive complexity results from an ILP formulation of the problem, by proving that the extremal optimal solutions of the continuous relaxation are integral if the preferences are single-peaked on a tree or a path. Finally, we have identified a new class of graphs for which the complexity of the recognition problem is polynomial, that of pseudotrees. 

While the generalization of single-peakedness to arbitrary graphs make it more plausible to learn some preference structures in applications, an interesting research direction for future works would be to formulate the recognition problem as the determination of a \emph{maximum likelihood} graph (while possibly imposing that the graph is an axis or a tree) on which the preferences are single-peaked. Put another way, one would relax the requirement of perfect compatibility of the graph with the observed preferences in order to facilitate structure learning on real preference data.
As shown in the numerical experiments, for an high preference heterogeneity, the graph can indeed become very dense when the number of voters increases (as is often the case in real applications), thus making essential to consider a more flexible view of single-peakedness.
Some interesting works in this direction have already been carried out (recognition of nearly structured preferences, e.g., \cite{bredereck2016there,cornaz2012bounded,erdelyi2017computational,faliszewski2014complexity}), but a lot remains to be done to make the approach fully operational. 





\bibliographystyle{plain}
\bibliography{comsoc}


\begin{contact}
Bruno Escoffier\\
Sorbonne Université, CNRS, 
LIP6, F-75005 and 
Institut
Universitaire de France\\
Paris, France \\
\email{bruno.escoffier@lip6.fr}
\end{contact}

\begin{contact}
Olivier Spanjaard\\
Sorbonne Université, CNRS, 
LIP6, F-75005 \\
Paris, France\\
\email{olivier.spanjaard@lip6.fr}
\end{contact}

\begin{contact}
Magdaléna Tydrichov\'a\\
Sorbonne Université, CNRS, 
LIP6, F-75005 \\
Paris, France\\
\email{magdalena.tydrichova@lip6.fr}
\end{contact}


\begin{appendix}
\section{Computation of the expected number of necessary edges} \label{sec:necessary_edges}

In practice,  it is computationally cumbersome to enumerate all permutations to compute $P(\{j,k\})$ according to Equation~\ref{eq:Pjk} (page~\pageref{eq:Pjk}). We present here another approach to compute efficiently this value. Note that the maximal distance between two permutations of length $m$ is $m(m-1)/2$. 
For each value $\delta \in \{1, \hdots , m(m-1)/2 \} $, let $N^m(\delta)$ denote the number of permutations $R$ such that $d(R,R_0) = \delta$, and $N^m_{\{j,k\}}(\delta)$ the number of permutations $R=(\pi(1),\ldots,\pi(m))$ such that $\{\pi(1),\pi(2)\}$ = $\{j, k\}$ and $d(R,R_0) = \delta$. Then, $P(\{j,k\})$ can be computed as 
\begin{equation} \label{eq:Pjk_poly}
P(\{j,k\}) = \frac{\sum\limits_{\delta = 0}^{m(m-1)/2}N^m_{\{j,k\}}(\delta)\exp{(-\theta \delta)}}{\psi(\theta)}.
\end{equation}
The value $N^m_{\{j,k\}}(\delta)$ can be computed as follows: 
\begin{itemize}
    \item Firstly, we define the permutation $R_1\!=\!(\pi_1(1),\ldots,\pi_1(m))$ such that 
    $\pi_1(1) = j, \pi_1(2) = k$ and for each pair $i,l$ of candidates different from $j$ and $k$, we have $$ \pi_1^{-1}(i) < \pi_1^{-1}(l) \text{ if and only if } \pi_0^{-1}(i) < \pi_0^{-1}(l).$$ 
    Similarly, we define the permutation $R_2\!=\!(\pi_2(1),\ldots,\pi_2(m))$ such that $\pi_2(1)\!=\!k, \pi_2(2)\!=\!j$ and for each pair $i,l$ of candidates different from $j$ and $k$, we have $$\pi_2^{-1}(i) < \pi_2^{-1}(l) \text{ if and only if } \pi_0^{-1}(i) < \pi_0^{-1}(l).$$ 
    We denote by $\delta_1$ (resp. $\delta_2$) the Kendall tau distance between $R_0$ and $R_1$ (resp. $R_2$).
    \item 
    As $R_1$ (resp. $R_2$) is between\footnote{A ranking $R'\!=\!(\pi'(1),\ldots,\pi'(m))$ is \emph{between} rankings $R\!=\!(\pi(1),\ldots,\pi(m))$ and $R''\!=\!(\pi''(1),\ldots,\pi''(m))$ if, for any pair $i,j$ of candidates, $\pi^{-1}(i)<\pi^{-1}(j)$ and $\pi''^{-1}(i)<\pi''^{-1}(j)$ implies that $\pi'^{-1}(i)<\pi'^{-1}(j)$.}  $R_0$ and any permutation $R$ such that $\pi(1)\!=\!j$ and $\pi(2)\!=\!k$ (resp. $\pi(1)\!=\!k$ and $\pi(2)\!=\!j$), we have $d(R_0,R)\!=\!d(R_0,R_1)+d(R_1,R)$ (resp. $d(R_0,R)\!=\!d(R_0,R_2)+d(R_2,R)$) because the Kendall tau distance satisfies the betweenness condition\footnote{A distance $d$ satisfies the betweenness condition if for all $R,R',R''$ such that $R'$ is between $R$ and $R''$ we have $d(R,R'')=d(R,R')+d(R',R'')$.} \cite{kemeny1959mathematics}. Consequently, the number of permutations $R$ inducing the necessary edge $\{j,k\}$ and such that $d(R_0,R)\!=\!\delta$ can be computed as 
    $$N^m_{\{j,k\}}(\delta) = N^{m-2}(\delta - \delta_1) + N^{m-2}(\delta - \delta_2)$$
    because $d(R_0,R_1)\!=\!\delta_1$ and $d(R_0,R_2)\!=\!\delta_2$ by definition of $\delta_1$ and $\delta_2$.  
    Note that this equation is well defined because $N^m(\delta)$ is fully characterized by $m$ and $\delta$. The problem consists now in determining the values $N^{m-2}(\delta - \delta_1)$ and $N^{m-2}(\delta - \delta_2)$. In this purpose, for any value $m$ and distance $\delta$, $N^m(\delta)$ can be computed thanks to the following recursion principle:
    \begin{itemize}
        \item Regardless of the length $m$ of the permutations considered, $N^m(0) = 1$ as there is only one permutation at distance 0 of $R_0$ - it is $R_0$ itself. 
        \item Let $m\!>\!1$ and $\delta\!>\!0$. Let $R$ be an arbitrary permutation of length $m$ at distance $\delta$ from $R_0$. The distance between $R_0$ and $R$ can be calculated as the number $\delta'$ of swap operations needed to move $\pi_0(1)$ to the first position in $R$, to which one adds the distance between the restrictions of $R_0$ and $R$ to their $(m-1)$ last elements. We have $0 \leq \delta' \leq m-1$, and so
        $N^m(\delta) = \sum_{\delta' = 0}^{\min(\delta, m-1)}N^{m-1}(\delta - \delta')$.
    \end{itemize}
    Overall, after a single preprocessing step in $O(m^4)$, each probability $P(\{j,k\})$ can be computed in $O(m^2)$. The preprocessing step consists in determining $N^i(\delta_i)$ for each $i \in \{1, \hdots , m\}$ and $\delta_i \in \{1, \hdots, i(i-1)/2\}$. Hence, there are $N(m)\!\in\!O(m^3)$ values to compute, and each of them is obtained in $O(m)$. Once these values are computed, Equation~\ref{eq:Pjk_poly} allows us to compute $P(\{j,k\})$ in $O(m^2)$. This is a significant improvement compared to the brute force implementation in $O(m!)$ of  Equation~\ref{eq:Pjk}.
\end{itemize}
\end{appendix}

\end{document}